# Modulating anomalous thermal quenching behavior of stimulation luminescence via high-orbit electronic satellite-stabilized Trap state in germanate-based phosphors for 5D optical data storage

*Wenqian Xu, Dangli Gao\*, Xiangyu Zhang, Wenna Gao, Dingjun Jia, and Yuhua Wang\**

W. Xu, D. Gao, W. Gao, D. Jia

College of Science, Xi'an University of Architecture and Technology, Xi'an, Shaanxi 710055, China

E-mail: gaodangli@xauat.edu.cn or gaodangli@163.com

X. Zhang

College of Science, Chang'an University, Xi'an, Shaanxi 710064, China

Y. Wang

National and Local Joint Engineering Laboratory for Optical Conversion Materials and Technology of National Development and Reform Commission, School of Materials and Energy, Lanzhou University, Lanzhou 730000, Gansu, China.

E-mail: wyh@lzu.edu.cn

**Abstract**

Persistent luminescence (PersL) materials, widely used in emergency lighting and information storage, are primarily employed at room temperature. However, their luminescent performance deteriorates sharply at high temperatures. Herein, a serials of $Mg_2GeO_4:Ti^{4+},Ln^{3+}$ (Ln = Tb, Eu) phosphors demonstrated anomalous thermal quenching PersL due to the temperature-dependent Fermi-Dirac distribution of bound charge carriers of $Ti^{4+}_{Mg^{2+}}$ as remote electron traps and $V_{Mg^{2+}}$ as hole traps. The high carrier retention rate of phosphors is attributed to the ability of $Ti^{4+}_{Mg^{2+}}$ positive charge center to strongly trap non-bonding electrons over a long range (about 20 Å) as the electronic satellite for its stable operation. Under external optical/thermal stimulation, the released electrons and holes recombine at the different luminescent levels of $Tb^{3+}$, resulting in the emission of different PersL branching ratios. Using these phosphors, we have developed 5D optical data storage (2D plane + trap depth + temperature + time) and the encrypted engine program for high-temperature aerospace engines. This study reveals the energy storage process of long-range trapping and releasing electrons by $Ti^{4+}$ electron traps, and provides a new design concept for the design of PersL materials.

# 1. Introduction

Energy storage and afterglow effects in persistent luminescence (PersL) materials have been widely explored during the past several decades due to their wide range of applications covering night-vision illumination, decorative landscaping, dosimetry systems, medical imaging, and anti-counterfeiting as well as information storage.[1-8] When a storage phosphor is irradiated by high-energy radiation (e.g., X-rays or ultraviolet light), part of the excitation energy is stored within the phosphor through the trapping of charge carriers (electrons or holes) in lattice defects or impurity sites.[9-12] This stored energy can be released via thermal, optical, or other physical stimulations, triggering stimulated emissions from luminescent centers in the material.[13] Storage phosphors have been categorized into two major types based on their energy release mechanisms including thermally stimulated storage phosphors and photostimulable storage phosphors.[14-16] Thermally stimulated storage phosphors release stored energy upon heating and photostimulable storage phosphors liberated energy through illumination with low-energy photons.[17,18]

Photostimulable storage phosphors and/or thermally stimulated storage phosphors with the photo/thermo stimulated (PSL/TSL) phenomenon serve ideal imaging plates for optical information storage and retrieval.[19-22] The information processing involves two distinct illumination steps (i.e., writing and reading). Exposure to X-rays or UV light "writes" a latent image by trapping electrons in the phosphor lattice.[23-25] The density of trapped electrons correlates with the local radiation dose, forming a temporary optical memory that remains readable (typically within 8 hours) post-exposure. Irradiation with low-energy photons/thermo triggers PSL/TSL,[26] releasing visible light proportional to the trapped charge density. Most electrons are discharged during readout, and residual traps are erased by bright fluorescent light, enabling plate reuse.[27,28] However, in the application of PersL materials for optical information storage, thermal disturbance from ambient temperature causes the uncontrollable loss of written information. The core reason lies in the metastable characteristics of information-storage traps.[29,30]

We must revisit the composition of afterglow materials: host lattice, luminescent centers, and trap centers. Both luminescent centers and trap centers rely on the host lattice and doping ion feature.[31] Thus, the structure of the host lattice, as well as the cationic properties and coordination environment of its constituent components, undoubtedly affect the luminescent centers and trap centers.[32-34] Afterglow is a type of localized state luminescence in solid materials, originating from the transition of localized electronic states within the band gap induced by lattice defects or doped ions in solids (**Figure** 1a). By nature, these localized-states are non-periodic and exhibit atomic-like energy levels within the bandgap, enabling them to act as effective trapping centers for free carriers or excitons. In most cases, the trapped electrons and holes typically undergo either radiative or non-radiative recombination, converting excitation energy into photons or phonons (Figure 1a), respectively.[35-37] Therefore, to enhance luminescence efficiency, the key lies in capturing excitation energy via localized-states with excellent luminescent properties, thereby reducing energy loss through other non-radiative recombination pathways. Figure 1a illustrates the competitive processes among various defect centers in solid materials, with three types of localized emission centers highlighted: self-trapped excitons (STE), emissive intrinsic defects, and intentionally doped ions.[38] How to transform non-radiative defect centers in materials into trap energy storage centers requires that the traps have comparable advantages to luminescent centers in the process of capturing free carriers or free excitons, and that the traps after capturing carriers have low formation energy — this is a major challenge in the material design process.

Figure 1b presents four main types of transition luminescence in intentionally doped materials. It is evident that luminescence processes are essentially determined by the strength of electron-phonon coupling, the energy level distribution of localized states, and the modes of lattice relaxation. To obtain high-performance PersL materials, the greatest challenge in material design lies in requiring that the defect density must be sufficiently high while not introducing additional non-radiative relaxation.

Herein, we first select the layered $Mg_2GeO_4$ (MGO) lattice as the host matrix. The goal is for the matrix to not only possess efficient photogenerated carrier capability but also for its layered structure to provide preferential carrier transport channels, reducing carrier energy loss caused by lattice phonon scattering.[39-41] Secondly, We chose $Tb^{3+}$ as the luminescent center because the 5d orbitals of $Tb^{3+}$ extend into the conduction band (CB), giving it an excellent ability to combine with carriers for PersL.[42] Finally, we selected $Ti^{4+}$ with $^3d_0$ empty orbitals as the co-dopant, aiming to regulate the host charge distribution by leveraging Ti's property of easily generating electron-hole pairs [43,44] (owing to photocatalytic activity of solid solution $TiO_2$ in $Mg_2GeO_4$) that readily combine with other defect centers, respectively. The goal of this design is to: trap free carriers or excitons in metastable trap energy-storage centers. This process is achieved by $Ti^{4+}$-mediated long-range confinement of electronic satellites or trapping of photogenerated electrons coupling with Boltzmann distribution of lattice phonon, thereby enhancing the degree of electron-hole overlap, guiding carrier pathways (the green arrow path in Figure 1b), and then increasing the radiative recombination rate, and inhibited the non-radiative relaxation channels (the red arrow path in Figure 1b).

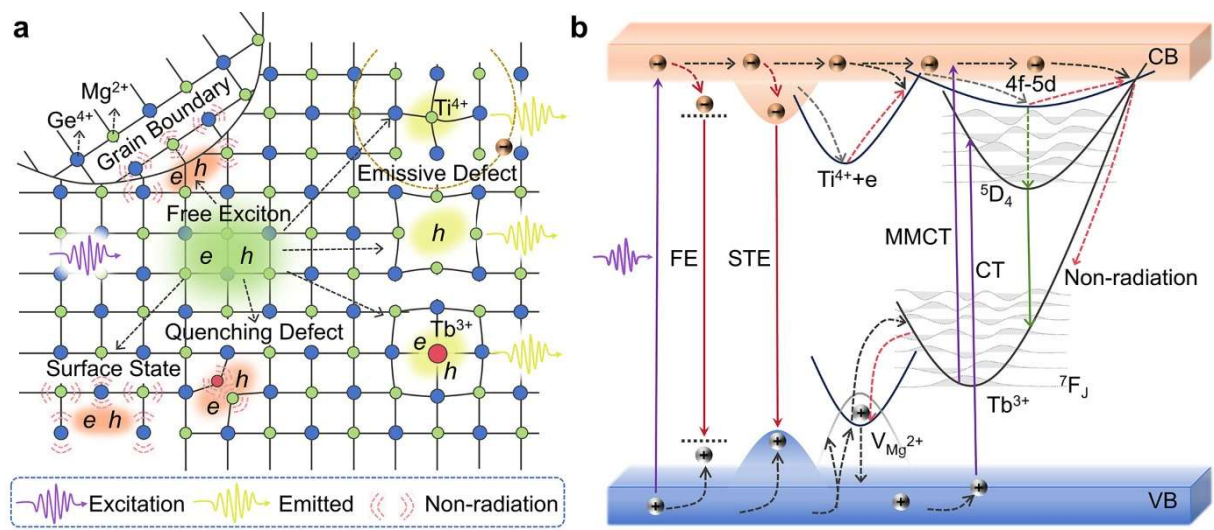

**Figure 1.** (a) Schematic illustrating main luminescent processes, including phonon-mediated non-radiative and radiative transition pathways, in semiconductor MGO:Ti,Tb materials driven by localized-state-induced lattice defects (i.e., composite potential traps including dislocation and grain boundary, electron or hole trap including interstitial, vacancy, antisite and doping). (b) Distribution of defect energy levels in localized states and related transitions (including free exciton (FE) emission, charge transfer (CT) transitions and metal-to-metal charge transfer (MMCT) transitions involve interactions between the energy levels of dopant ions and the band-edge states, and f-f/d transitions of dopant).

## 2. Results and discussion

### 2.1 Crystal structure

As shown in **Figure** S1a and Figure S2, the XRD diffraction peaks of all MGO-base samples match well with the standard card (PDF# 88-2303) of orthorhombic MGO, demonstrating that doping with $Ti^{4+}$, $Ln^{3+}$ (Ln = Tb, Sm, Dy, Pr and Eu), and $Li^+$ ions does not alter the host crystal structure. Rietveld refinement of the MGO matrix using GSAS software (Figure S1b) provided reliable structural parameters: $R$wp = 9.69%, $R$p = 7.35%, and $χ^2$ = 3.818, confirming the formation of pure-phase MGO. The orthorhombic crystal structure (space group *Pnma*) of MGO and the local coordination environments of $Mg^{2+}$ and $Ge^{4+}$ are illustrated in Figure S1c. One can find that parallel layers of lattice composed of $[Mg_2O_6]$ octahedra and $[GeO_4]$ tetrahedra are intercalated by a $[Mg_1O_6]$ octahedral layer.[40,41] Considering the effective ionic radii of $Mg^{2+}$ (r = 0.72 Å, CN = 6), $Ge^{4+}$ (r = 0.39 Å, CN = 4), $Ti^{4+}$ (r = 0.77 Å, CN = 6), $Tb^{3+}$ (r = 0.99 Å, CN = 6), and $Li^+$ (r = 0.76 Å, CN = 6), $Ti^{4+}$, $Tb^{3+}$, and $Li^+$ ions preferentially occupy $Mg^{2+}$ lattice sites.[42,45-47]

The upper panels of Figure S1d display the SEM image of MGO:Ti,Tb,Li, revealing irregular polyhedral particles with an average diameter of ~5.5 μm. Elemental mapping analyses (lower panels of Figure S1d) demonstrate the uniform distribution of Mg, Ge, O, Ti, and Tb elements throughout the sample, affording direct evidence of successful cationic incorporation into the MGO lattice. Notably, Li exhibits faint contrast in the mapping results, attributable to its low atomic mass and doping concentration.

**2.2 Luminescence performance of MGO tuned by co-doping Ti,Tb,Li**

**Figure** 2a presents the PL spectra ($\lambda_{ex}$ = 254 nm) and PLE spectra monitored at 556 nm in the MGO:Ti,Tb,Li, MGO:Ti,Tb and MGO:Tb phosphors. All samples exhibit analogous PL emission bands spanning 350-700 nm, assigned to the $^5D_{3,4} \rightarrow {}^7F_J$ $(J = 2-6)$ transitions of $Tb^{3+}$.[48] The PLE spectra monitoring at 556 nm feature broad-band peaking at 254 nm, originating from the 4f→5d orbital transitions of $Tb^{3+}$, accompanied by some characteristic sharp-line peaks corresponding to $Tb^{3+}$ $^7F_6 \rightarrow {}^5G_{2,3,4}$ in the 300-400 nm range.[42,49] Notably, neither PL nor PLE spectra of any sample exhibit distinct features attributable to $Ti^{4+}$, confirming that $Ti^{4+}$ ions serve exclusively as defect/trap centers rather than luminescent activators in MGO:Ti,Tb,Li and MGO:Ti,Tb phosphors.[50]

Figure 2b shows the PersL emission spectrum recorded at 5 s post-excitation, which mirrors the PL emission profile (Figure 2a), confirming that PersL originates from the $^5D_{3,4} \rightarrow {}^7F_J$ transitions of $Tb^{3+}$. Careful comparison reveals that while PersL spectra exhibits band positions and profiles consistent with PL spectra, they lack the fine spectral structures observed in PL (Figure 2a,b). This discrepancy arises from the distinct electron dynamics in the two processes: in PL, electrons undergo direct transitions between fluorescent energy levels, whereas PersL involves electron trapping, storage, and subsequent release via thermal or optical stimulation. Specifically, trapped electrons interact with optical or acoustic branch lattice waves; upon acquiring sufficient energy to surmount the trap barrier, they recombine with luminescent centers viaCB, emitting light. The absence of fine structure in PersL spectra indicates strong electron-phonon coupling between luminescent levels and matrix phonons, resulting in homogeneous linewidth broadening. This contrasts with the weaker electron-phonon interaction characteristic of PL processes.

Figure 2c presents PersL decay curves of the MGO:Ti,Tb,Li, MGO:Ti,Tb, and MGO:Tb phosphors after ceasing 254 nm UV excitation. Among these, MGO:Ti,Tb,Li exhibits the most superior PersL performance, with visible emission lasting up to 6 min in a darkroom. It is well established that PersL originates from the thermal-stimulated release of trapped carriers. To elucidate trap distribution and the PersL mechanism, thermoluminescence (TL) curves characterizing trap properties were measured (Figure 2d). For MGO:Ti,Tb,Li, MGO:Ti,Tb, and MGO:Tb samples, UV light charging was conducted for 6 min to investigate related trap dynamics. The TL curve profile of MGO:Ti,Tb,Li sample mirrors those of MGO:Tb counterpart, yet shows optimal TL intensity (Figure 2d), indicating that $Ti^{4+}$ and $Li^+$ co-doping enhances the overall trap concentration. Beyond intensity, the distribution of the TL curve reflects the characteristics of the traps. All three samples display broad TL responses spanning 300-650 K. It is worth noting that, when monitoring every peak PersL emission peaks of MGO:Ti,Tb,Li and MGO:Tb samples, all TL curves exhibit identical three-peak structures at 340 K (shallow traps), 439 K (moderate traps), and 565 K (deep traps). In contrast, MGO:Ti,Tb shows distinct TL features under the same emission monitoring, featuring a quasi-continuous broad band composed of four peaks at 346, 393, 445, and 565 K. The average trap depth ($E$) can be evaluated using the following equation:

$$E = (0.94 \ln \beta + 30.09) \times kT_m \tag{1}$$

where $E$, $\beta$, $k$, and $T_m$ represent the trap depth, heating rate, Boltzmann constant ($8.617 \times 10^{-5}$ eV/K), and the temperature of TL peaks [51]. The calculated trap depths corresponding to $T_1/T_2/T_3$ are about 0.882/1.138/1.465 eV in MGO:Ti,Tb,Li and MGO:Tb (Figure 2d). While trap depth is 0.897($T_1$)/1.019($T_0$)/1.154($T_2$)/1.465($T_3$) eV in MGO:Ti,Tb. These experimental results suggest that $Ti^{4+}$ doping introduces new traps (denoted as $T_0$), while $Li^+$ doping compensates for the non-isovalent doping effect of $Ti^{4+}$, but $Ti^{4+}$-$Li^+$ codoping together results in a high density of doping traps for information storage.

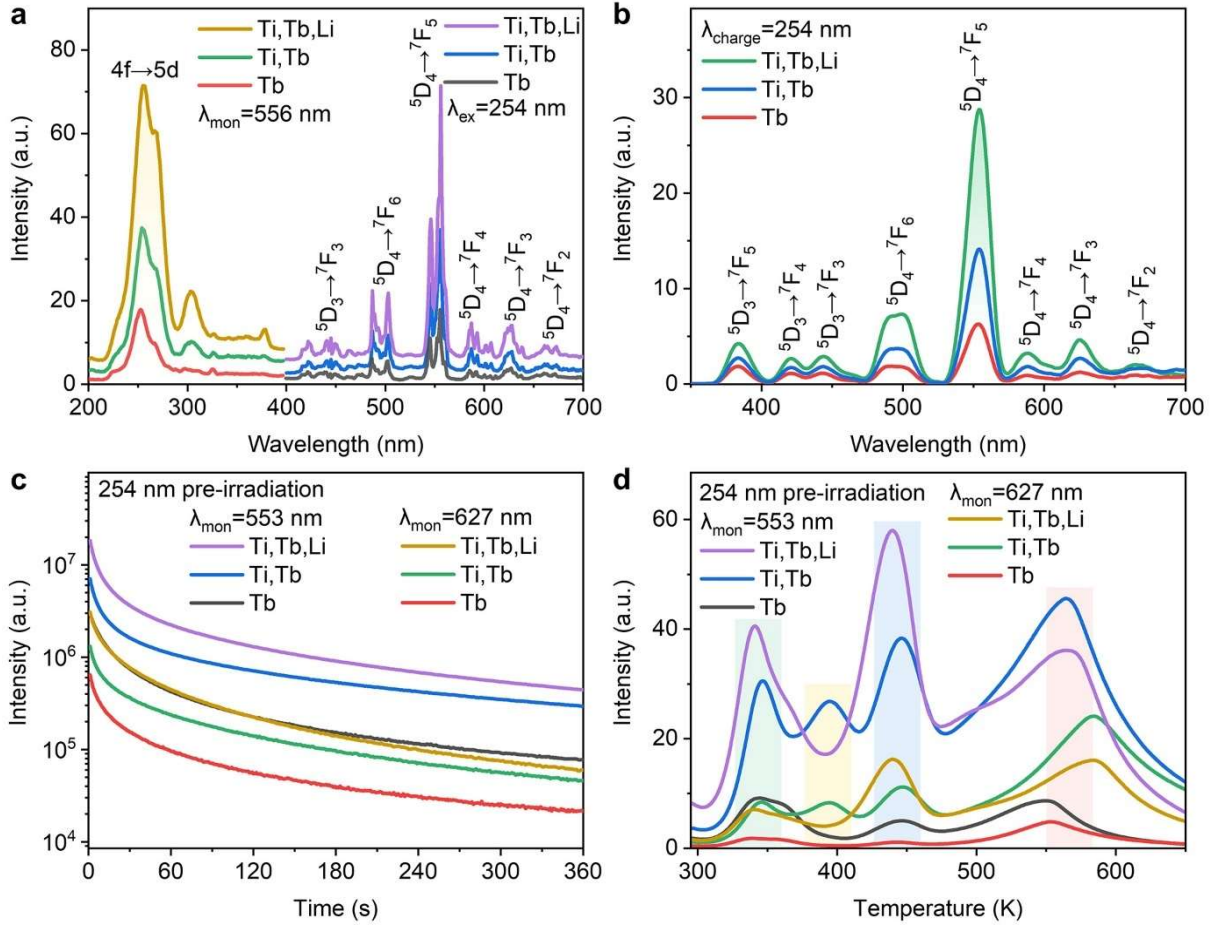

**Figure 2.** Luminescence characteristics of MGO-base phosphors including MGO:Ti,Tb,Li, MGO:Ti,Tb, and MGO:Tb. (a) PL and PLE spectra. (b) PersL spectra. (c) PersL decay curves. (d) TL curves. Notably, prior to measuring PersL, decay curves, and TL curves, the sample was annealed at 800 °C for 30 min to completely empty the traps, followed by pre-charging under 254 nm UV irradiation for 6 min.

To further confirm the matrix intrinsic traps and doping-introduced traps in MGO-based phosphors, a series of $Ln^{3+}$ (Ln = Eu, Pr, Dy and Sm) co-doped MGO:Ti samples were synthesized, and the corresponding spectral analyses are presented in Figure S3. We found that, similar to the MGO:Ti,Tb sample, all MGO:Ti,Ln (Ln = Eu, Pr, Dy, and Sm) samples exhibit the characteristic transitions of $Ln^{3+}$ and weak host transitions in their PL and PLE spectra (Figure S3a,b). Unlike the PL and PLE spectra of other samples, only MGO:Ti,Eu shows PersL emission spectral peaks of $Eu^{3+}$ at 615 nm, accompanied by a weak host broadband transition peak at 711 nm (Figure S3c). Monitoring the TL curves of different samples including MGO:Ti and MGO:Ti,Ln (Ln = Tb, Eu, Pr, Dy, and Sm), all samples at different PersL emission peaks reveal common TL peaks at approximately 350 K and 560 K, suggesting that these defect energy levels originate from host intrinsic defects such as $V_O$, $V_{Mg}$, and $V_{Ge}$ vacancies. The remaining traps can be attributed to doping-induced traps in MGO:Ti,Ln (Ln =

Tb, Eu, Pr, Dy, and Sm) samples (Figure 2d and Figure S3d). Among the five $Ti^{4+}$ and $Ln^{3+}$ co-doped samples, MGO:Ti,Tb exhibits the best PersL properties, with MGO:Ti,Eu being the next. The remaining three rare earth-doped samples do not show strong PersL properties. Therefore, in the following text, we focus on studying the luminescent properties of MGO:Ti,Tb, and MGO:Ti,Eu, MGO:Ti,Tb samples with Li doping for further charge balance regulation.

## 2.3 Temperature-dependent spectral characteristics of MGO:Ti,Tb,Li phosphors

To elucidate the trap-mediated electron trapping and release mechanisms in MGO-base phosphors, temperature-dependent PersL emission spectra and decay curves were systematically measured over a wide temperature range in three phosphors. As depicted in **Figure** 3a-d and Figure S4a,b, all three samples exhibit complex temperature-dependent behaviors in their PersL spectra. Specifically, MGO:Ti,Tb,Li displays two PersL intensity maxima at approximately 240 °C and 120 °C, whereas MGO:Ti,Tb exhibits maxima at around 240 °C and 110 °C (Figure 3a,b). These temperatures peak values align well with the peaks observed in their respective TL curves, indicating the presence of distinct trap energy barriers. For MGO:Tb (Figure S4), the first and second PersL intensity maxima occur at approximately 230 °C and room temperature (RT) within the investigated temperature range (Figure S4a). The temperature-dependent intensity variations across different spectral bands enable precise tuning of the PersL output color, as shown in the insets of Figure 3a,b and Figure S4a. Notably, when heated to 200-240 °C, all three samples exhibit a new, intense broadband emission spanning 550-750 nm (peaking at 627 nm), which is directly attributed to the $^5D_4 \rightarrow {}^7F_{2,3,4}$ transitions of $Tb^{3+}$.[48] These results suggest that the fluorescence branching ratio of intra-4f layer transitions depends on temperature.

Temperature-Dependent PersL Decay Kinetics (i.e., PersL decay curves) similarly exhibit complex temperature-dependent behaviors (Figure 3c,d and Figure S4b). For example, in MGO:Ti,Tb,Li, MGO:Ti,Tb, and MGO:Tb, the slowest decay rates coincide with the temperatures corresponding to the first and second maximum afterglow intensities (Figure 3c,d and Figure S4b). This behavior is rooted in the trap-mediated electron release mechanism. When excitation ceases, electrons trapped in shallow levels are rapidly liberated by room-temperature thermal activation, causing an initial sharp decline in afterglow intensity. As shallow trap charge carriers deplete, deep-trap electrons begin to release gradually, sustaining the afterglow at a slower decay rate. Notably, when the ambient temperature exceeds a trap's energy barrier (as denoted by TL curve peaks), charge carriers in adjacent deeper traps are activated, replenishing the emitting centers and decelerating the decay rate. This correlation is evident in the TL curve peaks, which typically correspond to the activation energies of matching traps. For example, in MGO:Ti,Tb,Li and MGO:Ti,Tb, the afterglow intensity declines rapidly within 30 s of excitation termination due to shallow-trap electron release. After 30 s, the decay transitions to a gentle regime dominated by deep-trap electron release (Figure 3c,d), illustrating how trap depth distribution dictates afterglow dynamics.

Both PersL intensity and duration exhibit complex thermal responses, while the PL intensity of all samples shows a monotonic decrease with increasing temperature (Figure S5a-c). This PL behavior is ascribed to the typical temperature quenching effect, which arises from thermal activation of multi-phonon non-radiative relaxation. The phenomenon can be quantitatively evaluated using the following formula:[52,53]

$$I_T = \frac{I_0}{1 + A\exp(-\Delta E/kT)} \qquad (2)$$

Here, $I_0$ and $I_T$ denote the emission intensities at the initial and test temperatures, respectively, with $k$ representing the Boltzmann constant and $A$ being a fitting constant. The linear fitting of $\ln(I_0/I_T-1)$ versus $1/(kT)$ is shown in Figure S6a. The fitting results reveal that the thermal quenching activation energies ($\Delta E$) of MGO:Ti,Tb,Li, MGO:Ti,Tb, and MGO:Tb phosphors are 0.254 eV, 0.178 eV, and 0.211 eV, respectively. As a result, MGO:Ti,Tb,Li has the largest thermal quenching activation energy, leading to the highest thermal stability. The CIE chromaticity coordinates of MGO:Ti,Tb,Li, MGO:Ti,Tb, and MGO:T b phosphors are presented in Figure S5d-f and Figure S6b, with detailed values listed in Table 1. CIE chromaticity coordinates nearly independent of temperature variations were obtained, indicating that the emission color remains stable over a wide temperature range.

Considering the combined temperature dependencies of PersL and PL intensities, we infer that the pronounced enhancement in PersL intensity for the $Tb^{3+}$ $^5D_4 \rightarrow {}^7F_{2,3,4}$ transition with increasing temperature is closely linked to the improved transition probability. This enhancement arises from the temperature-dependent Boltzmann distribution of hole states in the fine energy levels of the $Tb^{3+}$ ground state($f(E) \approx e^{-(E-E_F)/kT}$), where $E$ is the energy (eV), $E_F$ is the Fermi level, $k$ is the Boltzmann constant, and $T$ is the thermodynamic temperature) approximating the Fermi-Dirac distribution ($f(E)=1/(1+e^{-(E-E_F)/kT})$) at the $^7F_{2,3,4}$ energy levels (Figure 3a ,b).

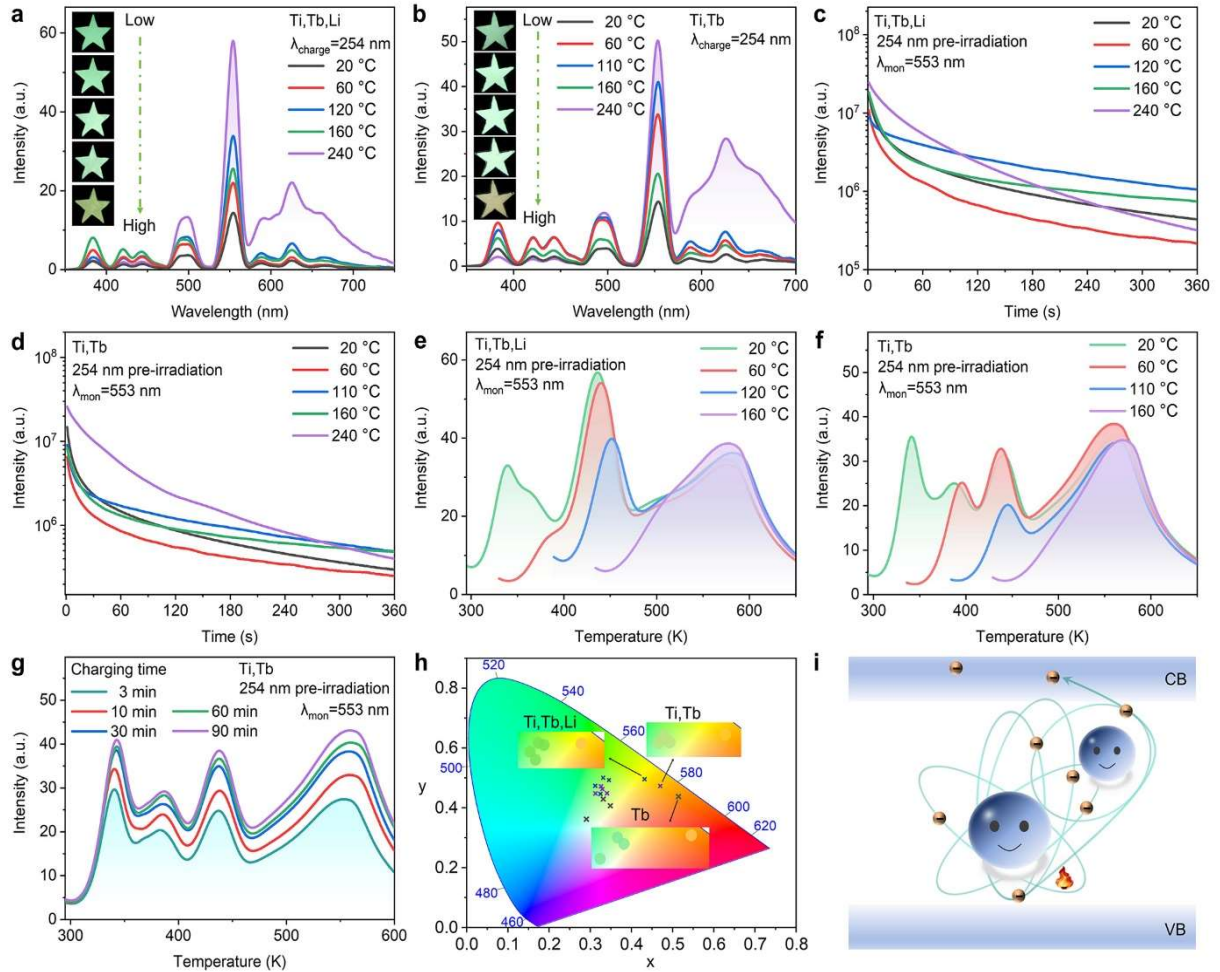

**Figure 3.** Temperature-dependent spectral characteristics of MGO:Ti,Tb,Li and MGO:Ti,Tb phosphors. (a,b) PersL spectra. (c,d) PersL decay curves. (e,f) TL curves. (g) TL curves as a function of charging time for MGO:Ti,Tb. (h) The chromaticity coordinates of the three PersL samples including MGO:Ti,Tb,Li, MGO:Ti,Tb and MGO:Tb after 254 nm UV charging. Insets show their photographs of PersL. (i) Schematic diagram of the detrapping process of defect-related satellite electrons under stimulation. All samples were pre-charged under 254 nm UV irradiation for 6 min prior to measurement.

To further reveal the temperature-dependent trap carrier population, TL curve partial-cleaning experiments were conducted according to McKeever's method in these samples.[54] The phosphor was first heated to a specific excitation temperature, charged under 254 nm UV light for 6 min, rapidly cooled to room temperature (RT), and then reheated to record the TL curve. A series of TL curves obtained at different charging temperatures are shown in Figure 3e,f and Figure S4c. Heating to 60-80 °C completely empties the shallowest trap, as evidenced by the first TL peak of MGO:Ti,Tb,Li, MGO:Ti,Tb, and MGO:Tb occurring at 60-80 °C. At 160 °C, charge carriers in the deepest traps of all samples are barely released, leading to maximum TL intensities at 230-240 °C (Figure 3e,f and Figure S4c). As charging temperature increases, low-temperature TL band intensities vanish first, followed by gradual diminishment of mid- and high-temperature bands.

Consequently, the overall profile of a set of temperature-dependent TL curves exhibits the disappearance of shallow-trap TL components alongside a slight shift of deep-trap components toward higher temperatures, indicating carrier redistribution. More specifically, as ambient temperature increases, trap carriers become unstable and redistribute. Fluorescent energy levels and deep-trap levels acquire higher carrier populations than RT, following the approximate Boltzmann distribution. The temperature-dependent TL intensities in Figure 3e,f demonstrate the dependence of fluorescent level populations on trap activation energy temperature, confirming their thermal dependency. Figure 3e,f show that TL curves shift slightly toward higher temperatures with increasing charging temperature. Specifically, shallow-trap carriers are emptied, while deep-trap carrier filling density slightly increases and new deep traps are filled. These experimental observations validate the concept of Boltzmann distribution for trapping energy level populations. Notably, the carriers in mid- and high-temperature TL bands remain unchanged (Figure 3e,f), providing a robust basis for trap-mediated RT carrier information storage. The temperature-modulated anomalous thermal quenching PersL effect can also be observed in MGO:Ti,Eu and MGO:Ti (Figure S7).

Considering optical information storage, the carrier storage capacity (i.e., trap filling capacity) represents a critical parameter. Take MGO:Ti,Tb as an example, the trap filling capacity reaches saturation (i.e., the maximum value of the TL integral intensity) after 90 minutes of UV charging, demonstrating an exceptional storage capacity comparable to that of an optical battery (Figure 3g). Notably, as the charging time prolongs, the filling capacity of deep traps increases more rapidly than that of shallow traps, which can be attributed to the photothermal effects. This effect enhances the interaction between optical-acoustic branch lattice wave and electrons, contributing to the observed phenomenon. As temperature increases, the redshift of PersL output color (Figure 3h) confirms that the temperature-related electronic state distribution regulates the emission spectrum profile. The abnormal thermal quenching phenomenon can be explained by the detrapping and migration of trapped satellite carriers under thermal stimulation, as illustrated in Figure 3i. As the temperature increases, the trapped satellite carriers absorb the host lattice vibration energy, migrate to CB after gaining sufficient thermal activation energy, and are finally captured by the luminescence centers to emit light. Furthermore, the number of escaped carriers is positively correlated with temperature (Figure 3i). The anomalous thermal quenching behavior of PersL in these phosphors enables controllable thermal/optical stimulation luminescence (TSL/PSL). As shown in Figure 4 and Figure S8, the pronounced TL enhancement under thermal stimulation across diverse temperature ranges demonstrates that charge carriers stored in traps of varying depths can be selectively extracted via a trap-addressing strategy through precise temperature management. Furthermore, these phosphors exhibit temperature-tunable on-demand TSL responses. Notably, MGO:Ti,Tb,Li and MGO:Ti,Tb exhibit the most significant TSL enhancement compared to their single-doped (e.g., MGO:Tb, MGO:Ti) and double-doped (e.g., MGO:Ti,Eu) counterparts (Figure 4a,b and Figure S8a-c). This superiority arises from the enhanced controllability of traps in MGO:Ti,Tb,Li and MGO:Ti,Tb across multiple thermal intervals, highlighting the critical role of co-doping in optimizing trap distribution for programmable luminescent applications. Besides the finely tuned PersL/TSL by heating, persL decay kinetics can also be modulated via photo-stimulation. As shown in Figure 4c,d and Figure S8d-f, pre-irradiated MGO:Ti,Tb,Li, MGO:Ti,Tb, MGO:Tb, MGO:Ti, and MGO:Ti,Eu phosphors exhibit rectangular pulse-like profiles in their decaying PersL curves under 808/980 nm laser diode illumination. These profiles consist of a markedly enhanced PSL (signal "1") and a drastically attenuated persistent stimulated photoluminescence (PSPL, signal "0").[55] The PSL intensity shows an inverse proportionality to the optical stimulation wavelength.

Detailed analysis reveals that under 980 nm stimulation, the PSL intensity undergoes a rapid decay during the first pulse stimulated period, followed by intensity stabilization in subsequent pulses—indicating rapid depletion of shallow-trap carriers in the initial pulse. In contrast, 808 nm pulsed laser excitation yields nearly constant PSL intensity across multiple pulses. Increasing 808 nm laser power enhances PSL intensity while maintaining stability, with no observable decay even after ~10 stimulation pulses (Figure 4c,d). This wavelength-dependent behavior highlights the potential for wavelength-selective trap addressing in programmable luminescent systems.

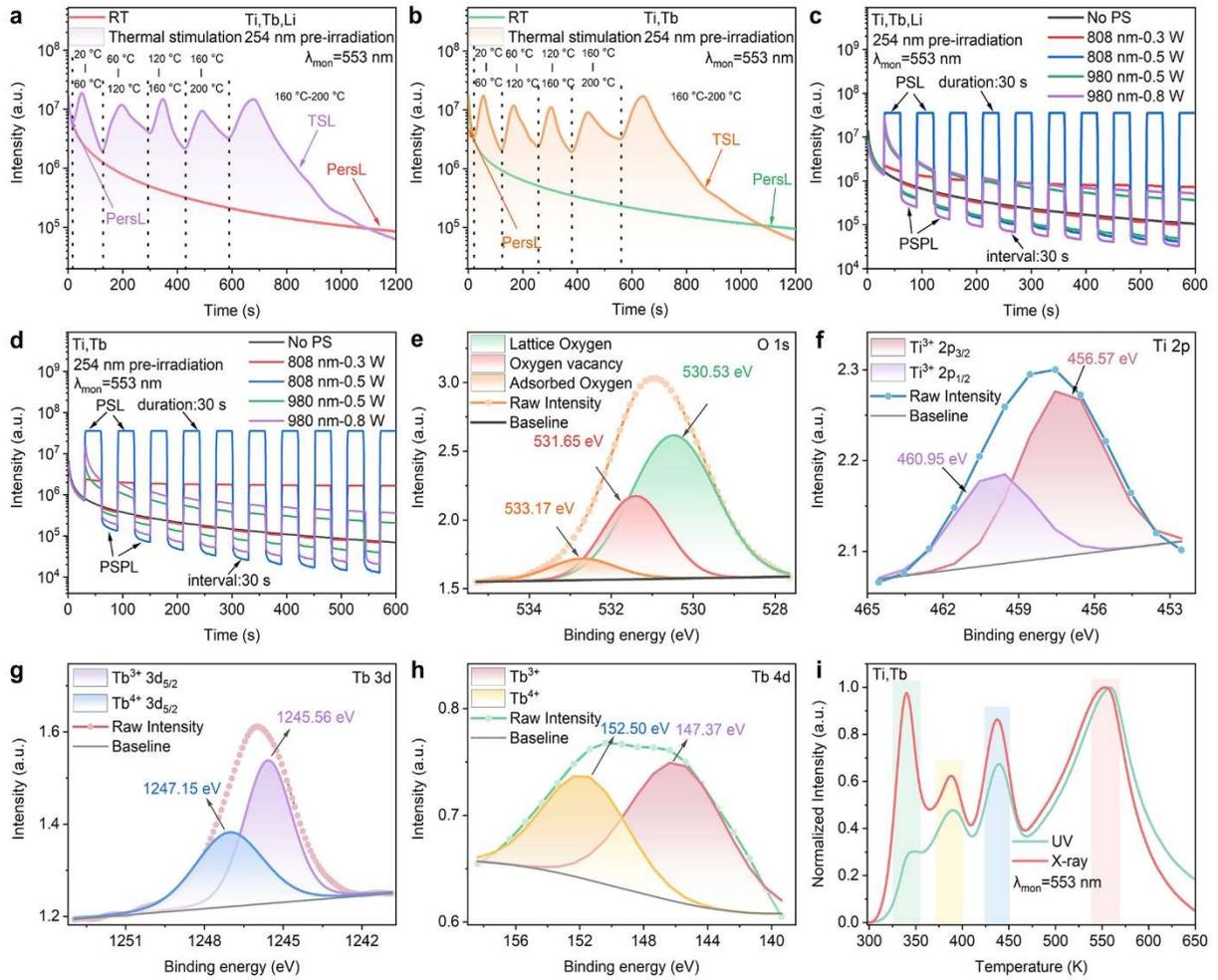

**Figure 4.** Stimulated luminescence characteristics of MGO:Ti,Tb,Li and MGO:Ti,Tb phosphors. (a,b) Temperature-controlled TSL on natural decay curves. (c,d) PSL modulated by pulse laser on natural decay curves. (e-h) high-resolution XPS spectra of O 1s, Ti 2p, Tb 3d and 4d in MGO:Ti,Tb phosphor. (i) The normalized TL curves charging by X-ray and 254 nm UV light. All samples were pre-irradiated with 254 nm UV light for 6 min prior to PersL decay and TL measurements.

Figure 4e-h show the high-resolution XPS spectra of the O 1s state in the host matrix, as well as the doped defect-related Ti 2p, and Tb 3d, 4d after X-ray irradiation. XPS spectra of O 1s clearly confirm that in the host lattice, there exist lattice oxygen (530.53 eV), oxygen vacancies (531.65 eV), and interstitial oxygen (533.17 eV) (Figure 4e), with an intensity ratio of approximately 4:2:1—indicating that the dominant intrinsic oxygen defects in the host matrix are oxygen vacancies rather than oxygen interstitials. Meanwhile, the high-resolution Ti 2p XPS spectrum is composed of 456.67 eV and 460.95 eV, and these binding energys are slightly less than that of $2p_{3/2}$ and $2p_{1/2}$ of $Ti^{3+}$. This suggests that after X-ray irradiation (the filling state of traps after X-ray irradiation is shown in Figure 4i, while the TL curves of other UV-charged samples are provided as references), $Ti^{4+}$ exhibits binding energies of $2p_{3/2}$ and $2p_{1/2}$ slightly

lower than those of $Ti^{3+}$, demonstrating that the $Ti^{4+}$ trapped 1–2 electrons. Note that both high-resolution XPS spectra of Tb $3d_{5/2}$ (1245.56 eV, 1247.15 eV) and $4d_{5/2}$ (147.37 eV, 152.50 eV) exhibits a double-peak characteristic, indicating the coexistence of $Tb^{3+}$ and $Tb^{3+}$+hole (Figure 4g,h).

To further confirm the carrier paths during PersL process, we measured charging wavelength-dependent TL curves and time-delayed PersL emission (Figure 5a-c). We found that except for differences in the relative intensities of the TL peaks, their positions are nearly independent of the excitation wavelength (Figure 5a,b). Additionally, the time-delayed emission spectra exhibited no changes in either their profiles or peak positions (Figure 5c). This observation suggests that after charging with different wavelengths, the energy levels of the luminescent centers remain unchanged, or in other words, the luminescent centers occupy a single lattice site. Furthermore, the correlation between TL curve morphologies and charging wavelengths indicates that most carriers recombine with luminescent centers via the conduction band, indicating a global PersL model.

Figure 5d demonstrates the time-delayed TL curves reveal that Trap II maintains an energy retention ratio of ~95% one hour after light-source shutdown. Even 10-30 days post-excitation, Trap II retains ~60% of its stored energy, after which the energy becomes nearly trapped indefinitely in deep-level traps, unable to escape at RT. To further reveal the depopulation approaches of trap carriers, we measured the TL curves of MGO:Ti,Tb phosphor after optical stimulation. Figure 5e-i compare the TL curve of the pre-charged MGO:Ti,Tb sample with and without optical stimulation, revealing that photo-stimulation is an inefficient method for depopulating traps, even when varying wavelength (down to 440 nm), irradiation time (60-600 s), or NIR laser power (up to 1.5 W). Neither shallow nor deep traps are significantly depleted under these conditions, explaining the negligible decay of PSL intensity after multiple 808 nm laser pulses—minimal changes in trap carrier concentration persist even after prolonged optical stimulation. In contrast, thermal stimulation effectively depopulates trapped carriers due to fundamental differences in excitation mechanisms. Thermal energy couples with acoustic branch lattice waves (characterized by continuous energy distributions and low group velocity), enabling efficient phonon absorption/emission without strict momentum/energy matching conditions. Conversely, stimulation light interacts with optical branch lattice waves, which exhibit high-frequency quantum behavior and require precise momentum-energy resonance for effective coupling. This stringent requirement often renders optical transitions transparent, limiting trap depopulation efficiency. These findings underscore the superiority of thermal over optical methods for manipulating trap carrier dynamics in these materials.

Compared to stimulation wavelength and irradiation time, stimulating light power exhibits relatively higher efficacy in releasing trap carriers. However, increasing NIR laser power introduces a non-negligible thermal effect, suggesting that carrier release may arise from thermal rather than optical processes. These results indicate that photo-stimulation not only fails to effectively retrieve information previously written by UV light but may also corrupt stored data via up-conversion effects.

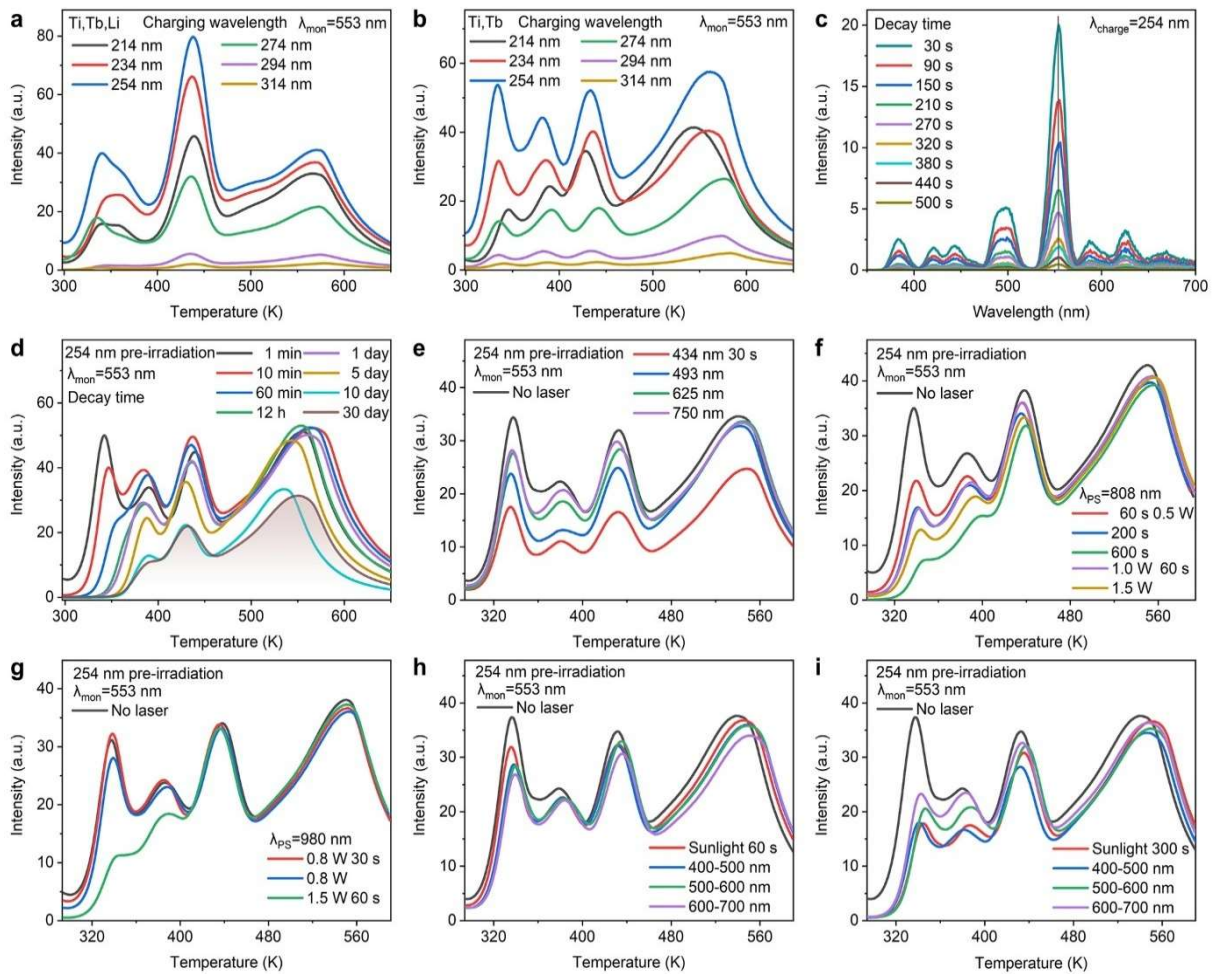

**Figure 5.** TL curves of MGO:Ti,Tb phosphor. (a,b) Charging wavelength-dependent TL curves. (c,d) Delay time-dependent PersL emission spectra and their TL curves. (e) TL curves after photo-stimulation with monochromatic xenon lamp at varying wavelengths. (f,g) TL curves following NIR laser stimulation with different powers or irradiation durations. (h,i) TL curves after stimulation of white light filtered by a band-pass filter. Prior to measurement, samples were annealed at 800 °C for 30 min to completely empty traps, followed by 6-min pre-charging under 254 nm UV irradiation.

To address the scientific question of whether the valence state of luminescent centers change during afterglow and whether they concurrently act as trap centers, we characterized the PL dynamics before and after charging. Figure S9 displays the PL dynamics of a series of MGO-based phosphors. Monitoring $Tb^{3+}$ green emission at 553 nm, no significant PL intensity variation was observed in MGO:Ti,Tb,Li, MGO:Ti,Tb, and MGO:Tb before/after charging—indicating that $Tb^{3+}$ ions undergo negligible valence change during afterglow (Figure S9a-c). Notably, in these samples, the $Tb^{3+}$ 4f→5d transition dominates over parity-forbidden 4f intra-configurational transitions. $Ti^{4+}$ doping enhances 4f→5d transition probability via lattice perturbation, leading to stronger 254 nm-excited PL in MGO:Ti,Tb,Li and MGO:Ti,Tb compared to MGO:Tb. The changes in the emission spectra and PLE spectra of the three groups of samples before and after charging are presented in Figure S10,11. By combining XPS (Figure 4f-h) and PL dynamic processes (Figure S9), it is plausible that the electrons bound by $Ti^{4+}$ originate from the vicinity of $Tb^{3+}$.

This lattice engineering strategy highlights the critical role of dopant-induced structural modulation in optimizing trap-luminescence coupling. $Ln^{3+}$ and $Ti^{4+}$ ions substitute for $Mg^{2+}$ in the MgO host lattice, inducing defects via the following defect reaction equation:

$$3Mg^{2+}+2Tb^{3+} \rightarrow 2Tb^{3+}{}_{Mg^{2+}}+V_{Mg^{2+}} \qquad (3)$$

$$2Mg^{2+}+Ti^{4+} \rightarrow Ti^{4+}{}_{Mg^{2+}}+V_{Mg^{2+}} \qquad (4)$$

$$3Mg^{2+}+2Li^{+}+Ti^{4+} \rightarrow 2Li^{+}{}_{Mg^{2+}}+Ti^{4+}{}_{Mg^{2+}} \qquad (5)$$

$$2Mg^{2+}+Tb^{3+}+Li^{+} \rightarrow Tb^{3+}{}_{Mg^{2+}}+Li^{+}{}_{Mg^{2+}} \qquad (6)$$

From defect equation (3), $Tb^{3+}$ ions substituting for $Mg^{2+}$ in MGO:Tb resides in a positive charge coordination environment, precluding oxidation to $Tb^{4+}$. Consequently, these $Tb^{3+}$ ions stably function as luminescent centers. The dominant defect centers are oxygen vacancies inherent to the oxide matrix and $Mg^{2+}$ vacancies introduced by doping. In MGO:Ti,Tb (defect equations 3,4), $Ti^{4+}$ substitution at $Mg^{2+}$ sites facilitates the trapping of electrons, whereas $Mg^{2+}$ vacancies are trapped by holes. This mechanism accounts for the increased trap density and modified TL curve profiles induced by $Ti^{4+}$ doping. For MGO:Ti,Tb,Li (defect equation 5), $Li^{+}$ doping at $Mg^{2+}$ sites functions as hole traps, reducing the concentration of $Mg^{2+}$ vacancy traps (defect equation 6). Notably, this substitution enhances material rigidity without compromising trap density, achieving a balance between structural stability and energy storage capacity.

The microscopic band-gap structure of electrons of MGO is shown in Figure S12. MGO demonstrates a direct bandgap of 3.27 eV, indicating a good optical matrix material. The PDOS analysis reveals that conduction band minimum (CBM) is primarily composed of Ge-4s and O-2p hybrid orbitals, while valence band maximum (VBM) is mainly contributed by O-2p orbitals. While $Ti^{4+}$ 3d exists at CBM, indicating that photogenerated electrons from O 2p VB electrons are easily trapped by $Ti^{4+}$ as as high-orbit satellite electrons.

## 2.4 PersL mechanisms

Spectral analysis, TL curves, XPS, band-gap structure of electrons, and defect equations collectively elucidate the regulatory mechanism of $Ti^{4+}/Li^+$ co-doping on the luminescence behavior of $Tb^{3+}$-activated phosphors (**Figure** 6a). Under UV irradiation, photo-generated electrons from VB or the ground state of $Tb^{3+}$ are elevated to CB, leaving holes in the VB. Photo-generated carriers migrate through the CB/VB and then are trapped by electron traps (e.g., $V_O^{\bullet\bullet}$, $Ti^{4+}{}_{Mg^{2+}}$) or hole traps (e.g., $V_{Mg^{2+}}$, $Li^+{}_{Mg^{2+}}$). After ceasing UV irradiation, trapping electrons/holes are released from traps to the CB/VB under thermal/optical stimulation, and then these electrons and holes recombine at the luminescent center of $Tb^{3+}$ and release energy in the form of light (i.e., PersL and TSL/PSL). In comparing thermal process (red curved dashed line stands for electron-phonon interaction) and photo-stimulation process (vertical black dashed line represents electron-photon interaction), thermal stimulation offers distinct advantages over optical methods for on-demand PersL regulation.

From the configuration coordinate diagram in Figure 6a, we can also observe that when carriers are populated into the ground state and excited state of the luminescent centers via photo-stimulation or thermal stimulation, the mutual Coulombic attraction suppresses the non-radiative relaxation pathways between the excited state and the ground state. During the entire closed-loop PersL process, the charge transfer pathways can be illustrated by the schematic diagram in Figure 6b. Under UV light irradiation, photogenerated electron-hole pairs are generated from the O 2p orbitals in VB. Subsequently, the electrons and holes are trapped via CB/VB and stored by $Ti^{4+}$ and $V_{Mg^{2+}}$ traps (magnesium vacancies with +2 charge state), respectively. Under photo/thermal excitation, these released electrons and holes are subsequently trapped by $Tb^{3+}$ ions and recombine to emit light. The Bohr radius of electrons trapped by $Ti^{4+}$ with the +2 charge can be calculated using the hydrogen-like atom single-electron approximation model, with the specific calculation formula as follows:

$$E_n = -\frac{m^* 4e^4}{8\varepsilon_0^2 \varepsilon_r^2 h^2 n^2} = -\frac{(m^*/m_0)}{\varepsilon_r^2} \cdot \frac{4*13.6}{n}(eV) \tag{7}$$

$$a^* = \frac{h^2 \varepsilon_r \varepsilon_0}{2\pi e^2 m_e^*} = 2*0.53 \cdot \frac{m_0}{m_e} \cdot \varepsilon_r (Å) \tag{8}$$

where $E_n$ is is the energy level, while h, ε, $m_0$, $m^*$, and $a^*$ are Planck's constant, permittivity, electron mass, effective electron mass, and, the Bohr radius, respectively. If we take $m^* = 0.5\ m_0$ and $\varepsilon_r = 10$, then $E_1$ can be estimated as 0.28 eV (this is of the same order of magnitude as the trap depth calculated from the experimental TL curve), and $a^* = 21.72$ Å. Obviously, this Bohr radius is much larger than the unit cell parameter. Therefore, this is a long-range Coulombic binding interaction as the centripetal force, where the electron orbits the $Ti^{4+}$ nucleus in a high-orbit, low-velocity circular motion with a radius of approximately 20 Å (the long-range electron is depicted in Figure 6b). The potential energy is lower CBM, so electrons are trapped in this potential energy valley and cannot migrate freely. The orbital radius larger than the lattice parameter confirms that $Ti^{4+}$ does not undergo valence change after charging, but instead binds long-range electrons. This also explains why the binding energies of Ti $2p_{3/2}$ and Ti $2p_{1/2}$ are even smaller than those of $Ti^{3+}$ (Figure 4f). Under optical/thermal excitation, they absorb energy to break free from the potential energy barrier, jump back to CB. Similarly, the hole escapes to VB. Subsequently, the electron-hole pairs recombine with $Tb^{3+}$, releasing the stored energy in the form of light. Of course, the electron trap center and the hole center may also orbit a common center in a manner similar to a binary or ternary star system, which further reduces the orbital radius and increases the trap depth. This model confirmed that the electrons or holes trapped by the traps are "long-range bound" rather than confined to the defect centers. In the afterglow process, the defects and bound carriers follow a long-range interaction model, where the defects do not undergo valence changes. Electrons are lost from the VBM O 2p, transition to the empty orbitals of Ti 3d and O 2p in CB, and are then bound by $Ti^{4+}$. Electrons in the ground state $Tb^{3+}$ jump back to VB to fill the holes, completing the energy storage and charging processes.

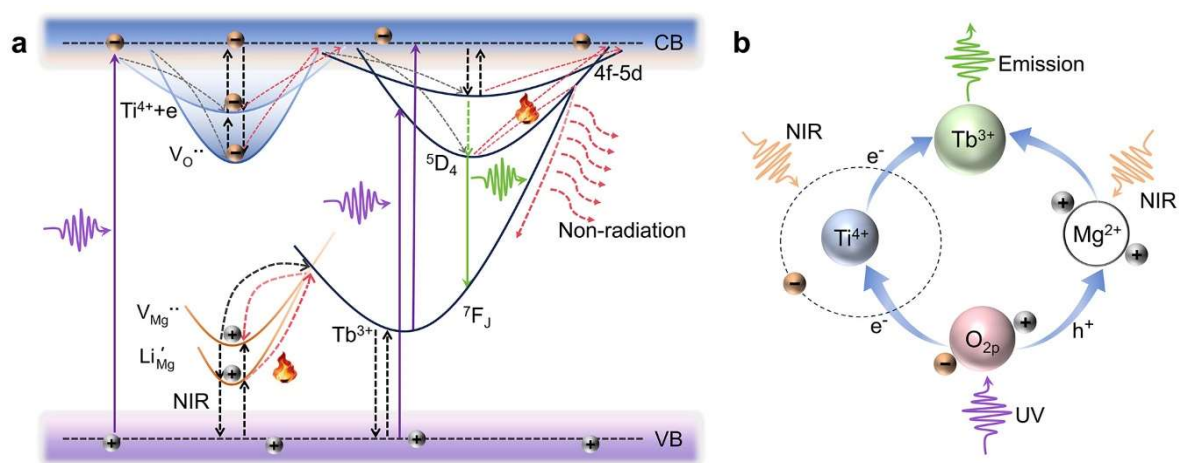

**Figure 6.** (a) Schematics illustrating the mechanisms of PL, PersL, TSL, PSL, and PSPL in the as-synthesized MGO:Ti,Tb,Li and MGO:Ti,Tb samples, with corresponding defect levels

marked in the diagram. Note that the purple and green vertical line represent excitation and emission processes, while vertical black dashed line and red curved dashed line stand for photon and thermo stimulation, respectively. (b) Charge transfer schematic diagram and electronic's long-range bound model by $Ti^{4+}$ in PersL Process.

**2.5 Dynamic luminescent patterns of MGO:Ti, MGO:Ti,Tb, MGO:Ti,Eu samples for five-dimensional optical data storage (two-dimensional plane + trap depth + temperature + time) and encryption**

Based on the storage characteristics of deep-trap carriers in these phosphors, we have designed a five-dimensional optical information storage system featuring "Fluorescent Feline's Dreamscape" using a single-layer phosphor film (**Figure** 7).[56-58] Figure 7a illustrates the writing and retrieval principles for five-level cartoon information of "Fluorescent Feline's Dreamscape" within a five-dimensional (5D) storage space (i.e., two-dimensional plane + trap depth + temperature + time) implemented in a real two-dimensional (2D) physical plane. Using 254 nm UV light irradiation through a patterned mask, distinct patterns corresponding to the five scenes of "Fluorescent Feline's Dreamscape"—including "Crescent Crouch", "Vine-Shadow Leap", "Lotus Drift", "Leaf-Wave Pounce", and "Vine-Weave Cocoon"—are written into traps at varying depths under different temperatures (Figure 7a). The filling capacity of traps at different depths is accurately controlled by managing charging time: 1 min, 2 min, 4 min, 7 min, and 10 min. As depicted in Figure 7b, "Vine-Weave Cocoon" and "Leaf-Wave Pounce" are initially written into the deepest traps (fourth and fifth levels) at 210 °C and 170 °C, respectively. This is followed by "Lotus Drift" and "Vine-Shadow Leap" transportation being written into the third and second traps at 120 °C and 70 °C, and finally "Crescent Crouch Sending" into the shallowest trap. These stored patterns can only be retrieved when the phosphor film receives adequate activation energy.[59] Leveraging the intelligent temperature response of the traps, sequential, individual, and distinct readout of the pre-stored "Fluorescent Feline's Dreamscape" scenes—"Crescent Crouch", "Vine-Shadow Leap", "Lotus Drift", "Leaf-Wave Pounce", and "Vine-Weave Cocoon"—is demonstrated in Figure 7b under thermostimulation at designated temperatures (60, 110, 160, 200, and 250 °C). These luminescent patterns are subsequently captured using a camera or mobile phone. Consequently, "Fluorescent Feline's Dreamscape" information can be stored and retrieved from the same location, enabling high-throughput optical data recording in five dimensions within a single recording layer.

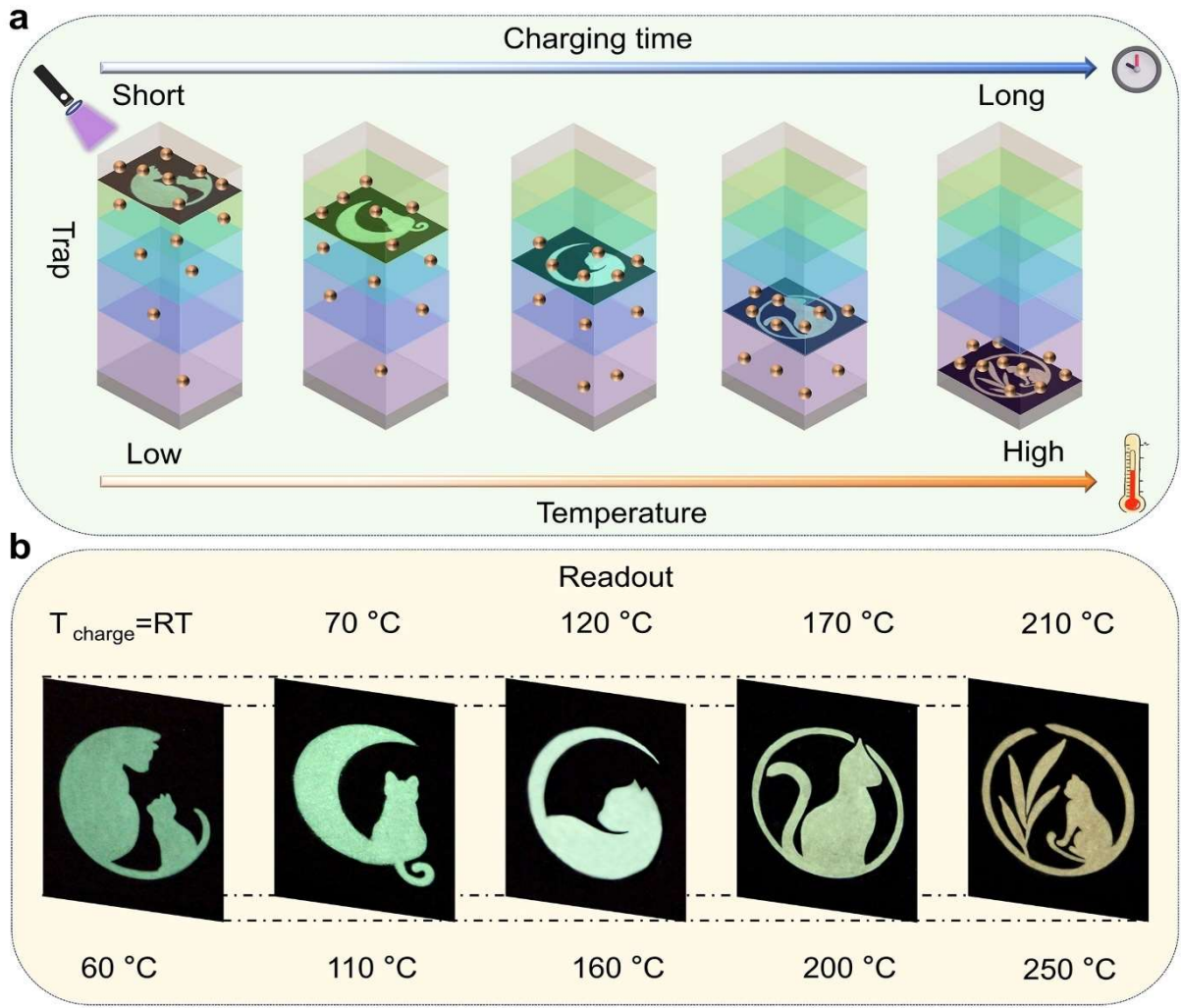

**Figure 7.** Five-dimensional optical data storage (two-dimensional plane + trap depth + temperature + time) and encryption of "Fluorescent Feline's Dreamscape" in single record layer of MGO:Ti,Tb flexible phosphor film with five multiplexed states including "Crescent Crouch", "Vine-Shadow Leap", "Lotus Drift", "Leaf-Wave Pounce" and "Vine-Weave Cocoon". (a) Schematic illustration of multilevel optical data storage writing via selective filling of traps, controlled by varying charging temperature and duration. Information is written using 254 nm UV light irradiation for 1 min, 2 min, 4 min,7 min and 10 min to fill shallow, medium, and deep traps. (b) Readout of pattern, s including "Crescent Crouch", "Vine-Shadow Leap", "Lotus Drift", "Leaf-Wave Pounce" and "Vine-Weave Cocoon", stored in five-layer traps with a suitable filter. stored in five levels of traps with a suitable filter. The first, second, third, fourth, and fifth images depict the readout of "Fluorescent Feline's Dreamscape".

Figure 8a presents the corresponding criteria for information writing and reading. Figure 8b depicts a fluorescent keyboard display panel printed using different phosphors, while Figure 8c demonstrates the process of key information writing and decryption. The operator pre-writes the five-level encrypted information (left panel) in advance following the sequence from high to low temperature (210 °C, 170 °C, 120 °C, 70 °C, RT) (left panel of Figure 8c). Subsequently, the aerospace engine begins to heat up and prepares to enter its operating mode. If the engine temperature sequentially rises from room temperature to 60 °C, 110 °C, 160 °C, 200 °C, and 250 °C, and the display panel sequentially displays the digital sequence shown in the right panel, the engine will activate the ignition mode and operate normally. Otherwise, it will automatically shut down for maintenance.

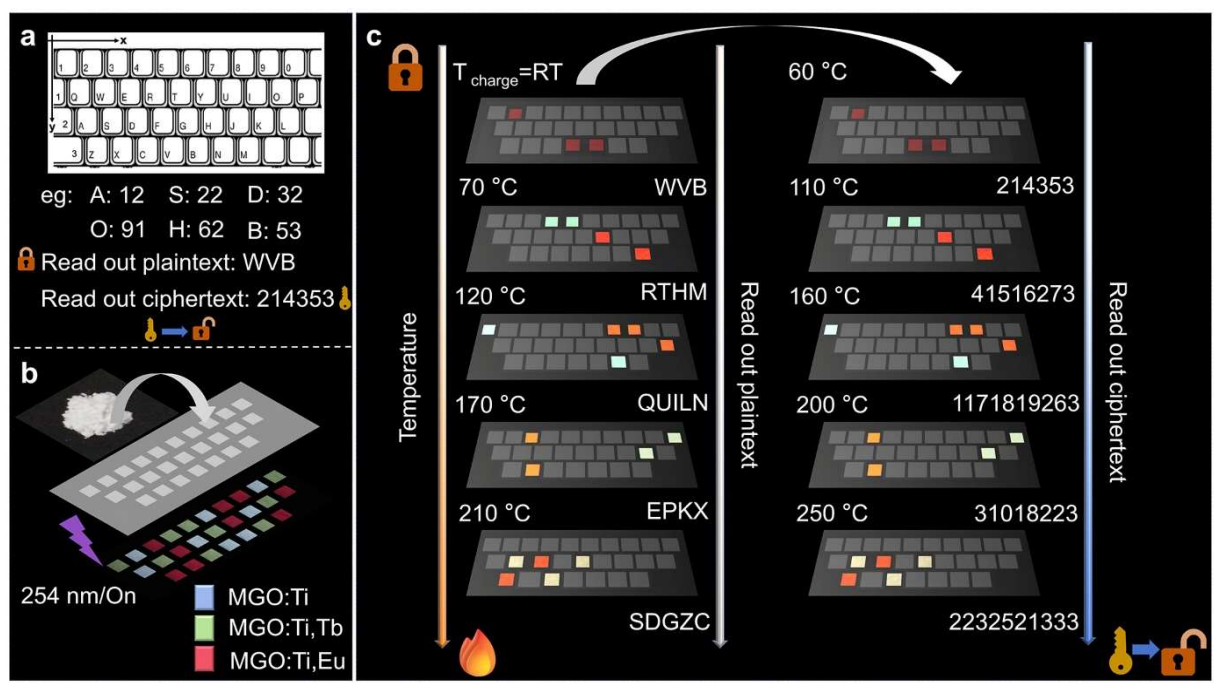

**Figure 8.** The key display panel for aerospace engines using MGO:Ti, MGO:Ti,Tb, MGO:Ti,Eu phosphors as the key for unlocking the engine program. (a) Diagram of the key display panel. (b) Blue, green and red regions are printed using MGO:Ti, MGO:Ti,Tb and MGO:Ti,Eu phosphor, respectively. (c) Schematic illustration of the principle for controlling the information writing and reading. When the temperature sequence of the read information (right panel) matches the reverse of the temperature sequence of the written information (left panel) based on the heating temperature, the correct password sequence is obtained to unlock the engine, which then automatically proceeds to the next program.

## 3. Conclusion

In summary, MGO:Ti,Ln (Ln = Tb, Eu) phosphors with ultra-broadband deep traps were successfully synthesized via a solid-state reaction. Co-doping with $Ti^{4+}$ ions significantly enhanced PersL performance of $Ln^{3+}$ activators. Notably, the introduction of $Li^+$ ions into MGO:Ti,Tb further improved the density of deep traps. High carrier retention rate of deep traps in these MGO-base phosphors enable the precisely temperature-controlled trap depth gradient to charge and discharge carriers. A Temperature-dependent carrier charging and discharging control mechanism of traps based on a Fermi-Dirac distribution law of boson is proposed. Leveraging the charging and discharging of these trapped carriers enables easily controllable functions, we developed the key display panel for aerospace engines using MGO:Ti, MGO:Ti,Tb, MGO:Ti,Eu phosphors as the key for unlocking the engine program. These findings provide a generalizable framework for designing multifunctional PersL/PSL/TSL materials, laying the groundwork for their emerging applications in X-ray imaging, biotherapy, anti-counterfeiting technologies, and advanced hierarchical information storage systems.

## 4. Experimental Section

*Materials*: High-purity chemicals, including MgO (A.R.), $GeO_2$ (99.99%), $TiO_2$ (99.99%), $Tb_4O_7$ (99.99%), $Sm_2O_3$ (99.99%), $Dy_2O_3$ (99.99%), $Pr_6O_{11}$ (99.9%), $Eu_2O_3$ (99.9%), and $Li_2CO_3$ (99.99%), were sourced from Aladdin (Shanghai, China). All reagents were used as received without any further purification.

*Sample preparation*: A series of MGO:0.1%$Ti^{4+}$, MGO:0.1%$Ti^{4+}$,0.1%$Ln^{3+}$ (Ln = Tb, Sm, Dy, Pr and Eu), MGO:0.1%$Ti^{4+}$,0.1%$Tb^{3+}$,0.1%$Li^+$ phosphors were synthesized using the conventional solid-state reaction method. Stoichiometric amounts of the starting materials (MgO, $GeO_2$, $TiO_2$, $Tb_4O_7$, $Sm_2O_3$, $Dy_2O_3$, $Pr_6O_{11}$, $Eu_2O_3$, $Li_2CO_3$) were thoroughly mixed, placed in a covered alundum crucible, and initially calcined in air at 900 °C for 1 h. The resulting product was then subjected to a secondary sintering process at 1250 °C for 5 h. After natural cooling to room temperature, the obtained samples were ground into fine powders for subsequent characterization.

*Computational details*: Density functional theory (DFT) simulations were conducted using the Vienna ab initio simulation package (VASP) with projector-augmented wave (PAW) potentials in periodic supercell configurations.[60] Electron exchange-correlation interactions were described employing the Perdew-Wang 91 (PW91) formulation within the generalized gradient approximation (GGA). Plane-wave basis set expansion was implemented with a 400 eV energy cutoff. The computational convergence criteria were differentially applied: electronic self-consistency was achieved with $10^{-5}$ eV energy threshold, while atomic relaxation followed the $10^{-2}$ eV/Å force criterion.

*Material Characterization*: The crystal structure of as-synthesized MGO:Ti,Tb,Li phosphors were characterized by X-ray diffraction (XRD) using a D/Max2550VBt/PC diffractometer with Cu Kα radiation of 1.5418 Å (40 kV, 40 mA). Morphological and particle size analyses were performed via scanning electron microscopy (SEM, ZEISS

Gemini 300), with energy-dispersive X-ray spectroscopy (EDX) integrated in the SEM system for chemical composition verification. X-ray photoelectron spectroscopy (XPS) spectra were acquired using a PHI VersaProbe 4 system. Photoluminescence (PL), photoluminescence excitation (PLE) spectra, persistent luminescence (PersL) decay curves, and photo/thermo-stimulated luminescence (PSL/TSL) emission spectra were recorded using a Horiba PTI Quanta Master 8000 spectrofluorometer equipped with a 150 W xenon lamp and an R928P photomultiplier tube (detection range: 240-1000 nm), enabling measurements of photo-stimulated persistent luminescence (PSPL). TL properties were evaluated using a custom-built TL measurement device (temperature range: 20-300 °C; heating rate: 1 °C/s), with a 1-min delay between UV charging and TL measurement. Near-infrared stimulation sources included a 0-5 W power-tunable 980 nm laser diode (spot size: 6 mm × 6 mm) and a 0-2 W power-tunable 808 nm laser diode (spot size: 9 mm × 4.5 mm). Luminescent images were captured by a Canon EOS 60D camera. Prior to measurements, phosphors were annealed in a muffle furnace at 800 °C for 20 min to eliminate pre-existing electron traps. Bandpass filters were used to suppress stray light interference during optical characterizations.

**Supporting Information**

Supporting Information is available from the Wiley Online Library or from the author.

**Acknowledgements**

This work was supported by the National Natural Science Foundation of China (11604253), Shaanxi Fundamental Science Research Project for Mathematics and Physics (23JSY003), and Natural Science Basis Research Plan of Shaanxi Province (2025JC-YBMS-745).

Received: ((will be filled in by the editorial staff))
Revised: ((will be filled in by the editorial staff))
Published online: ((will be filled in by the editorial staff))